\documentclass[aps,prd,onecolumn,nofootinbib]{revtex4-2}
\usepackage{float}

\usepackage{amsmath, amssymb, amsfonts}
\usepackage{graphicx} 
\usepackage{hyperref} 
\usepackage{tensor}   
\usepackage{physics}  
\usepackage[utf8]{inputenc}
\usepackage[english]{babel} 

\begin{document}

\title{Optical properties of gravitating strings}

\author{Marcos Silva}
\email{marcos.viniciussantos@ufpe.br}

\author{Azadeh Mohammadi}

\email{azadeh.mohammadi@ufpe.br}
\affiliation{Departamento de Física, Universidade Federal de Pernambuco, Recife, PE, Brazil}

\date{\today}

\begin{abstract}

We study the optical properties of gravitating Abelian-Higgs cosmic strings and compare them with those of the idealized infinitely thin string. We analyze the structure of the corresponding vortex solutions, characterizing their width, curvature profile, and approach to the ideal string limit. By investigating photon propagation in the string spacetime, we show that the finite core of the vortex gives rise to distinctive observational signatures absent in the ideal string approximation, including a characteristic triple-imaging configuration, strong demagnification of the central image, and a nontrivial Shapiro time delay between external and internal images. We determine how these effects depend on the parameters of the Abelian-Higgs model and show that the sign of the time delay is controlled by the ratio of the gauge boson mass to the Higgs boson mass, causing the string core to behave as either a temporal shortcut or a temporal barrier. Our results demonstrate that lensing effects can reveal information about the vortex formation and internal structure.

\end{abstract}

\maketitle

\section{Introduction}
\label{sec:intro}

Cosmic strings are cylindrically symmetric topological defects that arise in a wide class of field-theoretic models, typically through the spontaneous breaking of a local $U(1)$ symmetry. These objects are expected to have formed during highly energetic cosmological phase transitions in the early Universe \cite{kibble1976topology, kibble1980some, vilenkin1985cosmic}, such as those predicted by Grand Unified Theories (GUT). Mathematically, it can be shown that the predicted physical width of a GUT-scale string is negligible compared to its cosmological scales \cite{vilenkin1994cosmic}. Hence, when modeling their gravitational properties, it is standard to treat them with zero thickness. This approach is widely known as the ideal string model or the \emph{wire approximation}, in which the defect is simplified to a one-dimensional object with no internal degrees of freedom and characterized only by its linear mass density.

While cosmic strings at the GUT scale are extremely massive and characterized by short-range field interactions, making the ideal Nambu-Goto description highly accurate, recent developments have shifted astrophysical interest toward less energetic, extended vortex solutions. For instance, ultra-light scalar fields have emerged as prominent candidates for describing dark matter structures at both galactic \cite{ferreira2021ultra} and cosmological scales \cite{schive2014cosmic}. In ultra-light dark matter (ULDM) frameworks, such fields can alleviate small-scale controversies, such as the core-cusp problem, either through quantum pressure \cite{hu2000fuzzy} or via self-interactions introduced by a quartic potential \cite{desjacques2018impact}. Furthermore, networks of macroscopic scalar vortices have been proposed to actively shape the structure of dark matter halos \cite{brax20253d, rindler2012angular}. In addition to these astrophysical motivations, many ULDM models are extended by gauging the global U(1) symmetry, since global symmetries are not expected to remain exact in the presence of quantum gravity \cite{kamionkowski1992planck}. 
Since gauging such pure-scalar frameworks naturally leads to the Abelian-Higgs theory, exploring its parameter space beyond the ideal constraint is doubly justified. Consequently, a systematic study of the complete parameter space of the Abelian-Higgs model allows one to capture both the physical transition toward lower-energy vortex configurations and the local-gauge structures that are highly relevant to modern cosmological scenarios.

Historically, the main path to search for cosmic strings has been through their gravitational effects. In particular, the pioneering works of Vilenkin, Gott, and Hiscock \cite{vilenkin1981gravitational, gott1985gravitational, hiscock1985exact} showed that an ideal cosmic string produces two images of the same source, separated by an angle proportional to the string's linear mass density. Recently, this idealized framework has been extended to study the lensing properties of inclined configurations \cite{bulygin2023theory} and the emergence of wave-optical diffraction effects \cite{fernandez2017emergence}. However, the ideal string approximation treats the string as an infinitely thin source. While this simplification captures the large-scale gravitational effects of the defect, it introduces a curvature singularity at the string axis and eliminates any information about its internal structure. To understand how the internal structure of the string affects the propagation of light, one must consider a fully gravitating field-theoretic model with a finite core. Regular gravitating cosmic string solutions in the Abelian-Higgs model were first obtained by Garfinkle \cite{garfinkle1985general} and Linet \cite{linet1985static}, and later systematically classified in \cite{christensen1999complete,brihaye2000classical}. In these regular solutions, the finite width of the string, determined by the scalar and gauge field profiles, removes the conical singularity present in the ideal string description and modifies the trajectories of nearby particles and light rays \cite{hartmann2010geodesic,hartmann2011detection,hartmann2012geodesic}. Despite these advances, the impact of the string core on gravitational lensing has not yet been systematically explored across the full parameter space of the Abelian-Higgs model.

To bridge this gap, we go beyond the ideal string approximation and systematically investigate the optical properties of finite-width gravitating cosmic strings. More specifically, we ask two questions. What observable signatures distinguish a gravitating cosmic string from its idealized counterpart? And how do these signatures depend on the parameters of the underlying vortex model? To answer these questions, we employ the Abelian-Higgs framework as a representative model of a finite-width cosmic string and analyze the trajectories of photons propagating through its gravitational field. Our main goal here is to understand how the finite core of a realistic gravitating cosmic string affects observable lensing signatures and to establish a connection between the microscopic parameters of the vortex and macroscopic astronomical observables. 

This work is organized as follows. Section \ref{sec:background_solution} introduces the gravitating Abelian-Higgs string solutions and examines how their width, curvature, and domain of existence depend on the model parameters. Section \ref{sec:geodesics} develops the Hamiltonian description of photon propagation in the string spacetime and establishes the appropriate initial conditions. In Section \ref{sec:optical_properties_theory}, we present the lensing configuration and derive the lens equation. Section \ref{sec:results} is devoted to the optical properties of gravitating strings, including the formation of multiple images, image magnification, and Shapiro time delays. Finally, Section \ref{sec:conclusion} summarizes our main results and discusses their implications for the observational identification of finite-width cosmic strings.

\section{The Abelian-Higgs cosmic string}
\label{sec:background_solution}
The properties of gravitating Abelian-Higgs strings were first studied in detail in \cite{christensen1999complete} and later further clarified in \cite{brihaye2000classical}. In this work, we extend these numerical investigations and analyze the corresponding solutions and their implications in greater detail.
The gravitating Abelian-Higgs model describes a complex scalar field, $\phi$, coupled to a gauge field, $A_\mu$, through a local $U(1)$ symmetry, including a symmetry-breaking Higgs potential, $V(|\phi|)$. The matter fields are minimally coupled to gravity through the spacetime metric $g_{\mu\nu}$. The complete dynamics of the system is governed by the action
\begin{align}
S &= \int{d^4x \sqrt{-g} \mathcal{L_{\text{AH}}}} ,\\
\mathcal{L_{\text{AH}}} &= \frac{1}{2} D_\mu \phi D^\mu \phi^* - \frac{1}{4} F_{\mu \nu} F^{\mu \nu} - V(|\phi|) + \frac{1}{16 \pi G} R,
\end{align}
where $D_\mu$ is the gauge-covariant derivative,
$F_{\mu\nu}$ the field-strength tensor associated with the gauge field, $g$ the determinant of the metric,
$R$ the Ricci scalar, $G$ Newton's gravitational constant, and the Higgs potential 
\begin{equation}
V(|\phi|) = \frac{\lambda}{4}(\eta^2 - |\phi|^2) ^2. 
\end{equation}
To obtain the string solutions, we employ the following cylindrically symmetric ansatz for both the spacetime metric and the matter fields
\begin{align}
ds^2 &= -N(r)^2 dt^2 + dr^2 + L(r)^2 d\varphi^2 + N(r)^2 dz^2,
\label{eq:metric_ansatz} \\
\phi &= \eta f(r) e^{i n \varphi},\\
A_\mu dx^\mu &= \frac{1}{e} (n - P(r)) d\varphi,
\end{align}
where the functions $N, L, P,$ and $f$ depend only on the radial coordinate $r$, and the integer $n$ denotes the winding number characterizing the vorticity. Throughout this work, we restrict our analysis to strings with unit winding number.
We can rewrite the equations of motion in terms of the dimensionless radial coordinate $\tilde{r} = \sqrt{\lambda \eta^2}r$, together with the dimensionless constants $\alpha = e^2/\lambda$ and $\gamma = 8 \pi G \eta^2$. Consequently, the equations of motion take the form
\begin{align}
    \frac{(L N N')'}{N^2 L} &= \gamma \left( \frac{P'^2}{2 \alpha L^2} - \frac{1}{4}(1 - f^2)^2 \right) , \label{eq:field1} \\
    \frac{(N^2 L')'}{N^2 L} &= -\gamma \left( \frac{P'^2}{2 \alpha L^2} + \frac{P^2 f^2}{L^2} + \frac{1}{4}(1 - f^2)^2 \right) , \label{eq:field2} \\
    \frac{L}{N^2} \left( \frac{N^2 P'}{L} \right)' &= \alpha f^2 P , \label{eq:field3} \\
    \frac{(N^2 L f')'}{N^2 L} &= f(f^2 - 1) + f \frac{P^2}{L^2} , \label{eq:field4}
\end{align}
where the prime denotes differentiation with respect to the dimensionless coordinate $\tilde{r}$. Regularity of the spacetime geometry at the origin, together with the requirement that the matter fields approach their vacuum values asymptotically, imposes the following boundary conditions:
\begin{align}
    N(0) &= 1, \quad N'(0) = 0, \label{eq:bc_N} \\
    L(0) &= 0, \quad L'(0) = 1, \label{eq:bc_L} \\
    P(0) &= n, \quad P(\infty) = 0, \label{eq:bc_P} \\
    f(0) &= 0, \quad f(\infty) = 1, \label{eq:bc_f}
\end{align}
where the asymptotic conditions imposed on the matter fields, $P$ and $f$, correspond to the standard Nielsen-Olesen vortex boundary conditions.

As demonstrated in \cite{christensen1999complete}, the solutions to the gravitating Abelian-Higgs system can be classified into three distinct branches based on the asymptotic behavior of their metric functions: the String, Melvin, and Kasner branches. Among these, only the String branch consists of globally regular solutions, that is, configurations where the metric functions remain well-behaved and non-vanishing across the entire radial domain, $r \in [0, \infty)$. Conversely, the Melvin and Kasner branches represent closed solutions, meaning the metric functions either vanish or must be truncated at a finite radial coordinate. In this work, we focus primarily on the Cosmic String branch and its associated solutions. For a more detailed discussion of the Melvin and Kasner branches, as well as their physical interpretation, we refer the reader to Refs.~\cite{christensen1999complete, brihaye2000classical}.
\begin{figure}
    \centering
    \includegraphics[width=0.85\linewidth]{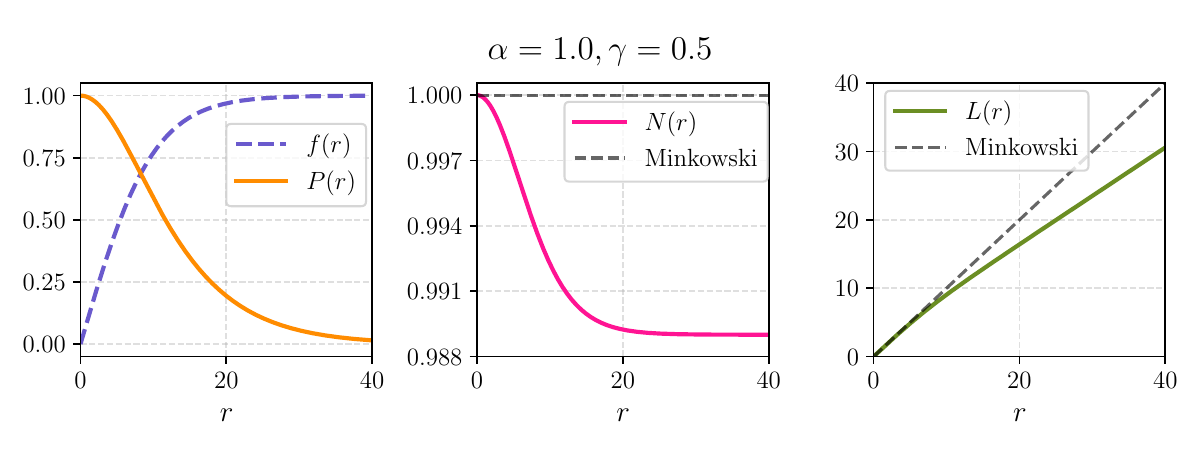}
    \caption{Radial profiles of the scalar and gauge fields (left), together with the metric functions
    $\sqrt{g_{tt}} = N(r)$ (center) and
    $\sqrt{g_{\varphi\varphi}} = L(r)$ (right).}
    \label{fig:field_profiles}
\end{figure}
Before discussing the properties of gravitating Abelian-Higgs strings in detail, it is useful to clarify the physical significance of the parameters $\alpha$ and $\gamma$. The parameter
\begin{align}
\alpha
&=
\frac{e^2}{\lambda}
=
\frac{1}{2}
\left(
\frac{m_A}{m_\phi}
\right)^2
=
\frac{1}{2}
\left(
\frac{\sqrt{2}\,e\eta}{\sqrt{\lambda}\,\eta}
\right)^2 ,
\end{align}
is proportional to the squared ratio between the vector boson mass and the Higgs boson mass. Physically, it controls the balance between two competing effects within the vortex core. The magnetic flux associated with the gauge field creates an outward repulsive force, while the scalar Higgs field exerts a restoring inward pull that confines the flux tube. 
It means that the parameter $\alpha$ controls the relative size of the gauge-field and Higgs-field masses and therefore determines the internal structure of the vortex core. Small values of $\alpha$ correspond to vortices with relatively broad magnetic cores, whereas large values of $\alpha$ favor stronger scalar-field localization and a more tightly confined flux tube. The critical value $\alpha=2$ corresponds to the self-dual or Bogomol’nyi limit of the Abelian-Higgs model. For $\alpha < 2$, the gauge-field penetration length exceeds the scalar coherence length, whereas for $\alpha > 2$ the scalar field becomes more strongly localized. Meanwhile, the parameter $\gamma \propto \eta^2$ dictates the energy scale of the spontaneous symmetry breaking.
 Throughout this work, we employ geometrized units, $c=\hbar=G=1$
and, without loss of generality, we set $\lambda = 1$, to facilitate the comparison of distinct vortex configurations.

Furthermore, as shown in Refs.~\cite{christensen1999complete,garfinkle1985general}, the metric functions $N(r)$ and $L(r)$ approach the following asymptotic functional form at large radial distances
\begin{equation}
    \begin{aligned}
        N(r \to \infty) &= a \\
        L(r \to \infty) &= br + c.
    \end{aligned}
    \label{eq:metric_asymp_limit}
\end{equation}
In Equation~\ref{eq:metric_asymp_limit}, the constants $a$, $b$, and $c$ all depend on the vortex parameters $\alpha$ and $\gamma$. The constant $b$ parametrizes the linear mass density of the string, $M$, which in the ideal string case satisfies the relation $b = 1 - 4M$. The asymptotic form of $L(r)$ describes a conical spacetime geometry with deficit angle $\delta\varphi = 2\pi(1 - b)$. The constant $a$ determines the asymptotic normalization of the temporal metric component and therefore measures the relative redshift between observers near the string core and observers in the asymptotic region.
Ref.~\cite{brihaye2000classical} showed that the asymptotic parameter satisfies $a<1$ when $\alpha<2$, whereas $a>1$ when $\alpha>2$. A particularly important case occurs at the critical coupling $\alpha = 2$. Similar to the flat-spacetime Nielsen-Olesen vortex~\cite{nielsen1973vortex}, it can be shown that the system admits first-order self-dual equations whose solutions automatically satisfy the full equations of motion,
Eqs.~(\ref{eq:field1})-(\ref{eq:field4}) \cite{comtet1988bogomol,linet1987vortex}. In this critical regime, the asymptotic parameters can be determined analytically and are given by
\begin{equation}
    a = 1, \quad b = 1 - \frac{\gamma}{2}, \quad \gamma_{\text{crit}}(2) = 2.
\end{equation}

\begin{figure}
    \centering
    \includegraphics[width=0.55\linewidth]{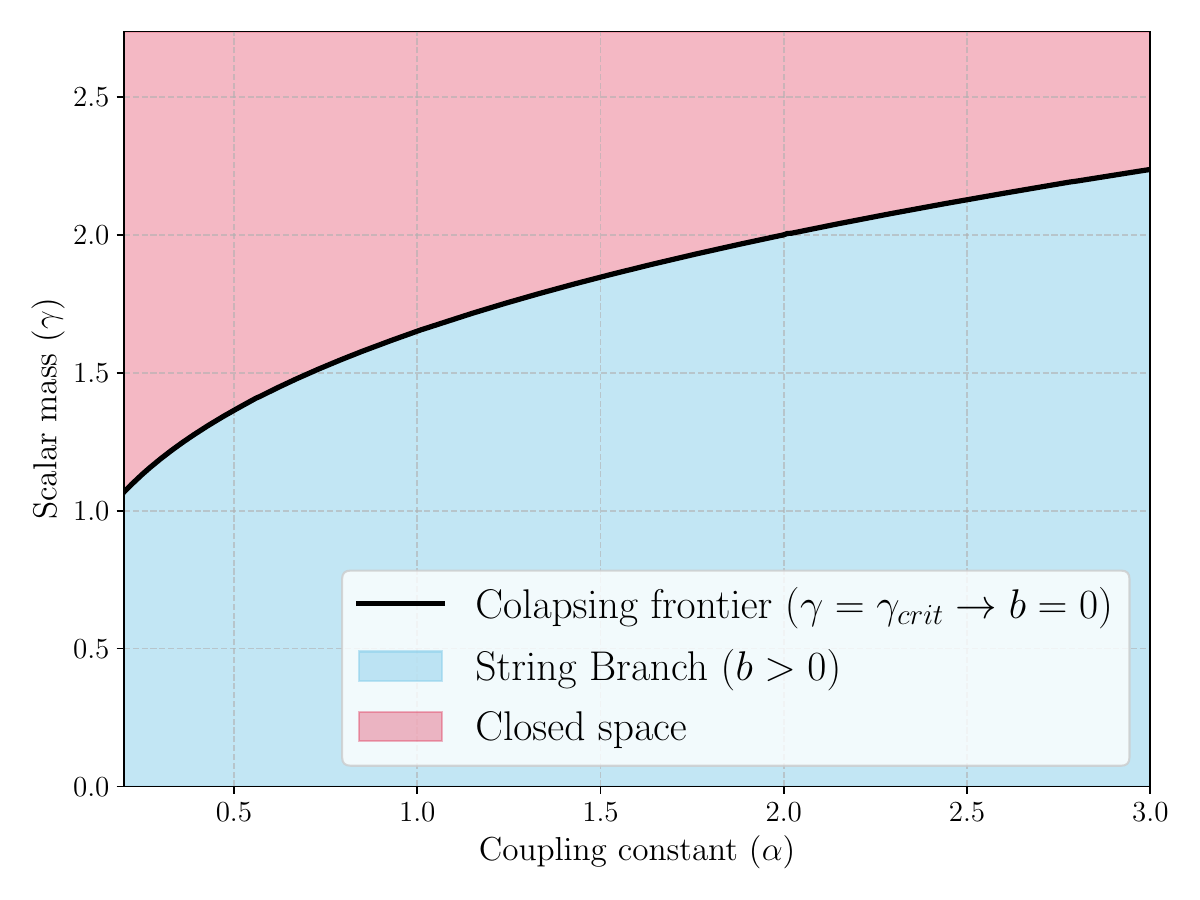}
    \caption{Critical curve $\gamma_{\mathrm{crit}}(\alpha)$ separating globally regular open string solutions from closed spacetime configurations. Notice that the allowed range of $\gamma$ increases monotonically with $\alpha$.}
    \label{fig:vortex_branch_diagram}
\end{figure}

Furthermore, for each fixed value of $\alpha$, there exists a critical coupling,
$\gamma = \gamma_{\mathrm{crit}}(\alpha)$, at which the asymptotic parameter satisfies $b(\alpha, \gamma_{\mathrm{crit}}) = 0$.
At this critical point, the conical deficit angle reaches its maximal value, $\delta\varphi = 2\pi$, defining the transition point between globally regular string solutions and closed spacetime configurations. Consequently, for $\gamma > \gamma_{\mathrm{crit}}(\alpha)$, the corresponding spacetime becomes closed. In Figure~\ref{fig:vortex_branch_diagram}, we present the critical curve $\gamma_{\mathrm{crit}}(\alpha)$ separating the regions of open and closed string solutions in parameter space.

\begin{figure}
    \centering
    \includegraphics[width=0.9\linewidth]{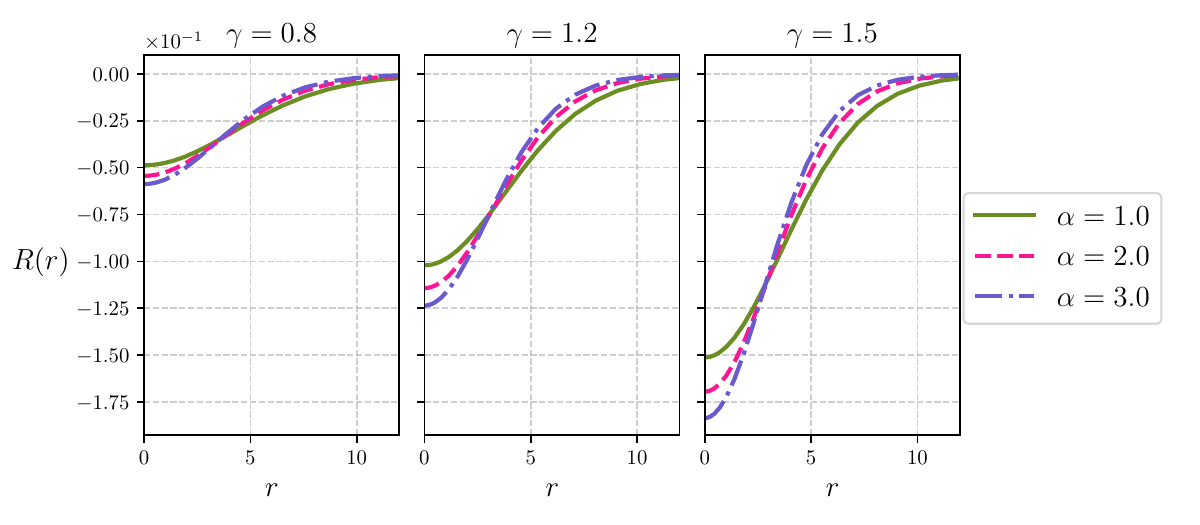}
    \caption{The curvature profile as a function of the coupling constant $\alpha$ and the energy scale of the string $\gamma$. }
    \label{fig:curvature_compare}
\end{figure}

The boundary conditions imposed on the metric function $L(r)$ eliminate the curvature singularity that is typically associated with the idealized thin-string description, which is commonly criticized (see Ref.~\cite{geroch1987strings} for example). Although the regularity conditions at the origin may initially suggest that the spacetime becomes locally flat near $r=0$, this is not generally the case, since the boundary conditions constrain only the values of the metric functions and their first derivatives, but not their higher-order derivatives.
In fact, plugging the boundary conditions into the equations of motion, Eqs.~\eqref{eq:field1}-\eqref{eq:field2}, shows that the second derivatives
$N''(0)$ and $L''(0)$ do not vanish identically. Consequently, the spacetime retains a nontrivial local curvature in the vicinity of the string core.
This can be seen explicitly from the Ricci scalar,
\begin{equation}
    R(r) = \frac{4 N''}{N} + \frac{2 (N')^2}{N^2} + \frac{2 L''}{L} + \frac{4 L' N'}{LN},
\end{equation}
which remains finite at the origin once the regularity conditions are imposed. In Figure~\ref{fig:curvature_compare}, we illustrate the dependence of the spacetime curvature on the parameters $(\alpha,\gamma)$. Notice that the vortex generates a smooth localized region of negative Ricci curvature around the origin, which gradually vanishes far from the string core. The symmetry-breaking scale, parametrized by $\gamma$, determines the magnitude of the curvature. For fixed $\gamma$, the curvature at the origin, $R(0)$, increases in magnitude with $\alpha$, meaning that stronger scalar-field confinement leads to a deeper curvature well around the string core.

\begin{figure}
    \centering
    \includegraphics[width=0.8\linewidth]{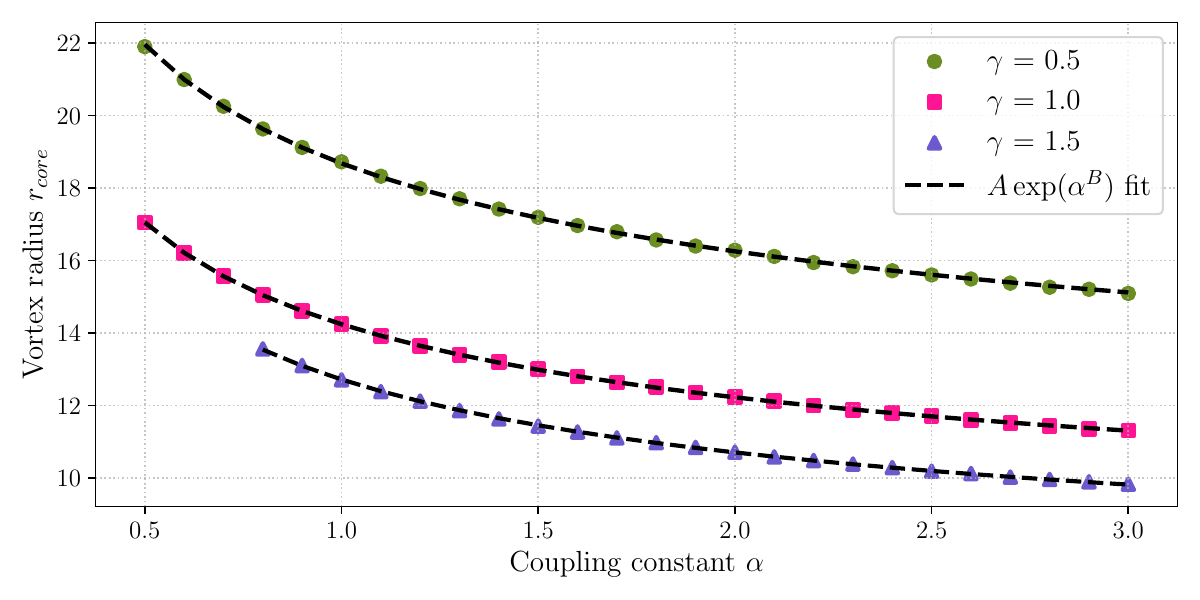}
    \caption{Vortex radius as a function of $\alpha$ for several energy scales $\gamma$. Increasing either the gauge coupling parameter $\alpha$ or the gravitational coupling $\gamma$ leads to a stronger localization of the vortex core. Missing data points around $\alpha = 0.5, \gamma = 1.5$ is due to $\gamma_{\text{critical}}(\alpha \approx 0.75) \leq 1.5$. }
    \label{fig:vortex_radii_compare}
\end{figure}

To characterize the curvature profile, we use two quantities: the central curvature, $R(0)$, and the vortex radius, $r_{\mathrm{core}}$, defined as $|R(r_{\text{core}})| = 0.01|R(0)|$. 
In Figure \ref{fig:vortex_radii_compare} we show how the string width changes with $\alpha, \gamma$. We observe that the string width decreases monotonically as either $\alpha$ or $\gamma$ increases. In particular, as $\alpha$ increases past $\alpha = 2$, the scalar-field confining effects become increasingly dominant, leading to a stronger localization of the vortex core and consequently to a narrower curvature profile.

\begin{figure}[ht]
    \centering
    \includegraphics[width=0.84\linewidth]{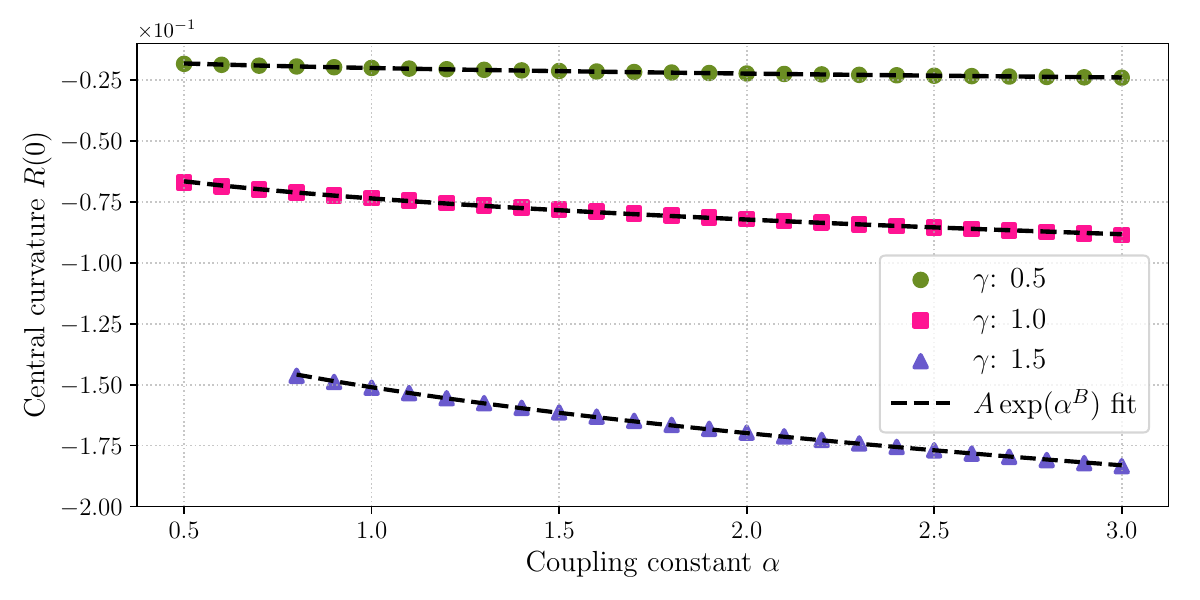}
    \caption{String central curvature $R(0)$, as a function of the parameters $\alpha$ and $\gamma$. The magnitude of the curvature well increases with both the confinement parameter $\alpha$ and the gravitational coupling $\gamma$.}
    \label{fig:vortex_centralcurv_compare}
\end{figure}

In Figure~\ref{fig:vortex_centralcurv_compare}, we observe that the absolute value of the Ricci curvature at the string core, $|R(0)|$, increases with the gravitational coupling parameter $\gamma$. This effect becomes increasingly pronounced for strongly confined vortex configurations, i.e., in the regime of large $\alpha$, where the scalar-field contribution dominates, and the vortex core becomes more localized.
We also see that the influence of $\alpha$ on the central curvature becomes stronger as the symmetry-breaking scale increases. In other words, gravitational effects amplify the sensitivity of the spacetime curvature to the internal structure of the vortex. These results suggest that the idealized zero-width cosmic string can be understood as a limiting configuration of the gravitating Abelian-Higgs vortex in the limit $\alpha \to \infty$ and $\gamma \to \infty$.

Interestingly, we find that both the vortex radius, $r_{\mathrm{core}}$, and the central curvature, $R(0)$, are remarkably well described by a power-exponential fitting function of the form $f(\alpha) = A \exp(\alpha^B)$, where $A$ and $B$ are fitting parameters that depend on the value of $\gamma$. 
For both quantities, this functional form reproduces the numerical data with high accuracy across the explored parameter range. It is worth mentioning that our choice of the fitting ansatz is based on two criteria: minimizing the number of free parameters and minimizing the mean-squared error.

\section{Geodesic Equations in Hamiltonian formalism}
\label{sec:geodesics}
In this section, we describe the mathematical procedure to obtain the geodesic equation in the spacetime generated by a gravitating Abelian-Higgs string.

\subsection{Cartesian coordinates}
Although the string solutions are naturally expressed in cylindrical coordinates, the coordinate system becomes singular at the origin ($r = 0$). To avoid numerical difficulties associated with this coordinate singularity, we reformulate the problem in Cartesian coordinates,
\begin{align}
x = r\cos\varphi, \quad
y = r\sin\varphi.
\end{align}
where $r=\sqrt{x^2+y^2}$. Differentiating these relations and inverting the transformation, we obtain
\begin{align}
    d\varphi = \frac{ x dy - y dx}{r^2}, \quad
    dr = \frac{x dx + y dy}{r}.
    \label{eq:dx_cylindrical}
\end{align}

Now, Substituting Eq.~\eqref{eq:dx_cylindrical} into the metric \eqref{eq:metric_ansatz}, we obtain the gravitating string metric in Cartesian coordinates,
\begin{equation}
    ds^2 = -N(r)^2 dt^2 + \frac{1}{r^2} \left(x^2 + \frac{L(r)^2 y^2}{r^2} \right)dx^2 + \frac{1}{r^2} \left(y^2 + \frac{L(r)^2 x^2}{r^2} \right)dy^2 + \frac{xy}{r^2} \left(1 - \frac{L(r)^2}{r^2} \right) dx dy.
    \label{eq:metric_cartesian}
\end{equation}
As expected, Eq.~\eqref{eq:metric_cartesian} reduces to the Minkowski metric when $N(r)=1$ and $L(r)=r$.
Note that, for the ideal cosmic string spacetime,
\begin{equation}
N(r)=1,
\qquad
L(r)=br.
\end{equation}
In this case, the mixed component $g_{xy}$ remains non-vanishing whenever ($b\neq1$). Thus, the conical geometry of the ideal string manifests itself through an off-diagonal metric component in Cartesian coordinates.

\subsection{Hamiltonian formalism}

The motion of a test particle in a curved spacetime can be described within the Hamiltonian formalism. The Hamiltonian associated with a metric $g_{\mu\nu}$ is
\begin{equation}
\mathcal{H}
=
\frac12 g^{\mu\nu} p_\mu p_\nu
=
-\frac{m^2}{2},
\label{eq:hamiltonian_general}
\end{equation}
where the second equality corresponds to the mass-shell condition. Restricting the motion to the plane perpendicular to the string ($z=\mathrm{const.}$) and using the metric~\eqref{eq:metric_ansatz}, Eq.~\eqref{eq:hamiltonian_general} becomes
\begin{equation}
    \mathcal{H} = \frac{1}{2} \left( -\frac{p^2_t}{N(r)^2} + p^2_r + \frac{p^2_\varphi}{L(r)^2} \right).
    \label{eq:hamiltonian}
\end{equation}
Since the geodesic equations will be integrated in Cartesian coordinates, we rewrite the Hamiltonian in terms of the canonical variables $(x,y,p_x,p_y)$. Under a coordinate transformation, the covariant components of the momentum transform according to
\begin{equation}
p_{\mu'}
=
\frac{\partial x^\nu}{\partial x^{\mu'}}\,p_\nu.
\end{equation}
Applying this relation to the transformation between cylindrical and Cartesian coordinates yields
\begin{align}
p_x
&=
\frac{x}{r}p_r
-
\frac{y}{r^2}p_\varphi,
\\
p_y
&=
\frac{y}{r}p_r
+
\frac{x}{r^2}p_\varphi.
\end{align}
Inverting these relations, we obtain
\begin{equation}
p_r
=
\frac{xp_x+yp_y}{r},
\qquad
p_\varphi
=
xp_y-yp_x.
\end{equation}
The second expression corresponds to the canonical angular momentum about the string axis.

Substituting these relations into Eq.~\eqref{eq:hamiltonian}, the Hamiltonian can be written entirely in Cartesian coordinates as
\begin{equation}
\mathcal H
=
\frac12
\left[
-\frac{p_t^2}{N(r)^2}
+
\left(
\frac{xp_x+yp_y}{r}
\right)^2
+
\frac{
\left(
xp_y-yp_x
\right)^2
}{L(r)^2}
\right].
\label{eq:hamiltonian_final}
\end{equation}
Since the metric is independent of $t$, the conjugate momentum $p_t$ is conserved along the geodesic. With the present sign convention, the conserved energy is $E=-p_t$.

The particle trajectory is then obtained by integrating Hamilton's equations,

\begin{equation}
\begin{aligned}
    \frac{\partial \mathcal{H}}{\partial x} &= -\dot{p}_x, \quad \frac{\partial \mathcal{H}}{\partial p_x} = \dot{x} \\
    \frac{\partial \mathcal{H}}{\partial y} &= -\dot{p}_y, \quad \frac{\partial \mathcal{H}}{\partial p_y} = \dot{y},
\end{aligned}
    \label{eq:hamilton_eqs}
\end{equation}
where the dot denotes differentiation with respect to the affine parameter.

The problem is therefore reduced to the numerical integration of four coupled first-order ordinary differential equations. Details of the numerical implementation are presented in Appendix~\ref{appendix:numerical appendix}.

\subsection{Initial conditions}

To integrate Eq.~\eqref{eq:hamilton_eqs}, it is necessary to specify initial conditions for both the particle position and the canonical momenta, $(x_0,y_0,(p_x)_0,(p_y)_0)$.
We consider light rays incident from the asymptotic region and initially propagating parallel to the $x$-axis. Consequently, $p^y = 0$.
This condition allows us to determine the initial values of the covariant momentum components $p_x$ and $p_y$.
Using the Cartesian metric~\eqref{eq:metric_cartesian}, the covariant and contravariant momentum components are related by
\begin{align}
    p_x &= g_{xx} p^x + g_{xy} p^y \xrightarrow{p^y = 0} g_{xx} p^x \\
    p_y &= g_{yy} p^y + g_{yx} p^x \xrightarrow{p^y = 0} g_{yx} p^x.
\end{align}

Since we are considering light rays, the four-momentum satisfies the null condition
\begin{equation}
g^{\mu\nu} p_\mu p_\nu
= 0.
\end{equation}

Using Eq.~\eqref{eq:metric_cartesian} together with $p^y=0$, we obtain
\begin{equation}
-\frac{p_t^2}{N(r)^2}
+
g_{xx}(p^x)^2
=
0.
\end{equation}

Since the conserved energy is given by $E=-p_t$, the contravariant component $p^x$ becomes
\begin{equation}
p^x
=
\frac{E}{N(r)\sqrt{g_{xx}}}.
\end{equation}

Substituting this result into the expressions for $p_x$ and $p_y$, we obtain the initial values of the covariant momentum components,
\begin{equation}
\begin{aligned}
(p_x)_0
&=
\frac{E}{N(r)}
\sqrt{g_{xx}},
\\
(p_y)_0
&=
\frac{E}{N(r)}
\frac{g_{xy}}{\sqrt{g_{xx}}}.
\end{aligned}
\label{eq:initial_condition_momentum}
\end{equation}

Together with the initial position $(x_0,y_0)$, Eq.~\eqref{eq:initial_condition_momentum} completely specifies the initial conditions required to integrate Eq.~\eqref{eq:hamilton_eqs}.
For each trajectory, the deflection angle is extracted from the asymptotic momentum components according to
\begin{equation}
\Theta(\xi)
=
\arctan\!\left(
\frac{p^y}{p^x}
\right),
\end{equation}
where the momentum components are evaluated at the final point of the trajectory.

\section{Optical properties: theory}
\label{sec:optical_properties_theory}

\begin{figure}[H]
    \centering
    \includegraphics[width=0.75\linewidth]{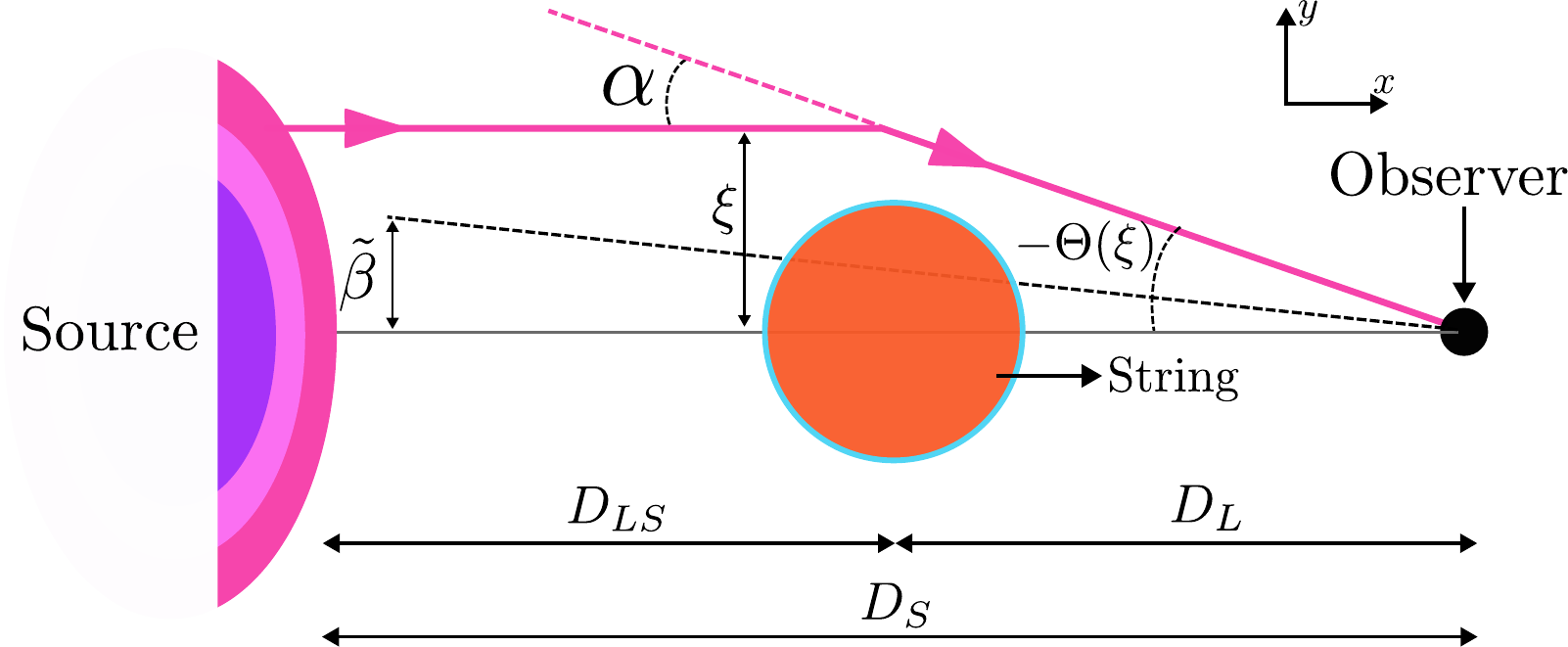}
    \caption{Top view of the cosmological setting.}
    \label{fig:lensing_setting}
\end{figure}

The lensing configuration in the cosmological setting we are studying here is shown in Figure~\ref{fig:lensing_setting}. A cosmic string lies between a distant light source and an observer, with the relevant distances shown in the figure. In this work, we adopt the thin-screen approximation, where all characteristic distances $D_i$ are assumed to be much larger than the size of the lensing region, allowing the interaction between the light rays and the string to be treated on a single lens plane. We consider a source located at a distance $D_S$ from the observer. Regardless of the specific form of the angular distribution at emission, since $ D_S \to \infty$, these rays reach the string as a bundle of parallel rays propagating along the $x$-direction with an impact parameter $\xi$. After interacting with the string spacetime, each ray is deflected by an angle $\Theta(\xi)$ and subsequently propagates toward the observer. Numerically, the deflection function $\Theta(\xi)$ is obtained by integrating a family of geodesics with different impact parameters. The trajectory with $\xi=0$ passes through the center of the string and experiences no net deflection, whereas trajectories with $\xi\neq0$ are bent toward the string core. 
By analyzing the resulting mapping between $\xi$ and $\Theta$, we find that multiple trajectories can connect the same source and observer. In fact, for any non-singular bounded transparent gravitational lens the number of images produced is shown to be odd \cite{burke1981multiple, mckenzie1985gravitational}. In particular, we identify a regime in which three distinct light rays reach the observer from the same source. This phenomenon, which does not occur for ideal cosmic string due to its singularity, will be referred to as \emph{triple imaging}. The existence of multiple images can be understood quantitatively through the lens equation. In particular, triple imaging arises whenever three distinct values of the impact parameter $\xi$ correspond to the same apparent source position. In what follows, we derive the corresponding lens equation and use it to characterize the image multiplicity produced by gravitating Abelian-Higgs strings.

The lens equation relates the angular position of the source in the absence of lensing (true source position), denoted by $\tilde{\beta}$, the angular position of the observed image (apparent image position), denoted by $\Theta$, and the deflection angle of the light ray due to the string, denoted by $\alpha$~\cite{meneghetti2021introduction}. In the thin-screen approximation, this relation is
\begin{equation}
    \Theta D_S
    =
    \tilde{\beta} D_S
    +
    \alpha D_{LS}.
    \label{eq:lens_equation_raw}
\end{equation}

We define the projected source position on the lens plane as $\beta=D_L \tilde{\beta}$
where $D_L$ is the observer-lens distance. For small angles, the impact parameter is
\begin{equation}
    \xi
    =
    D_L \tan\Theta
    \simeq
    D_L \Theta.
\end{equation}
Using these definitions, Eq.~\eqref{eq:lens_equation_raw} becomes
\begin{equation}
    \beta =
    \xi-
    D_{\mathrm{eff}} \alpha(\xi),
    \qquad
    D_{\mathrm{eff}}
    =
    \frac{D_L D_{LS}}{D_S}.
    \label{eq:lens_equation_alpha}
\end{equation}

The numerical integration of the geodesic equations gives the deflection
angle $\Theta(\xi)$. With our sign convention,
\begin{equation}
    \Theta(\xi)
    =
    -\alpha(\xi).
\end{equation}
Therefore, the lens equation can be written as
\begin{equation}
    \beta
    =
    \xi
    +
    D_{\mathrm{eff}}\Theta(\xi).
    \label{eq:lens_equation}
\end{equation}

In what follows, we use $\beta$ to characterize the image multiplicity.
This is equivalent to using $\tilde{\beta}$, since the two quantities are
related by the linear rescaling $\beta=D_L\tilde{\beta}$. Throughout this
work, we set $D_{\mathrm{eff}}=100$.

\section{Optical properties: Results}
\label{sec:results}

\subsection{Triple-imaging and angular separation}
\begin{figure}{h}
    \centering
    \includegraphics[width=0.7\linewidth]{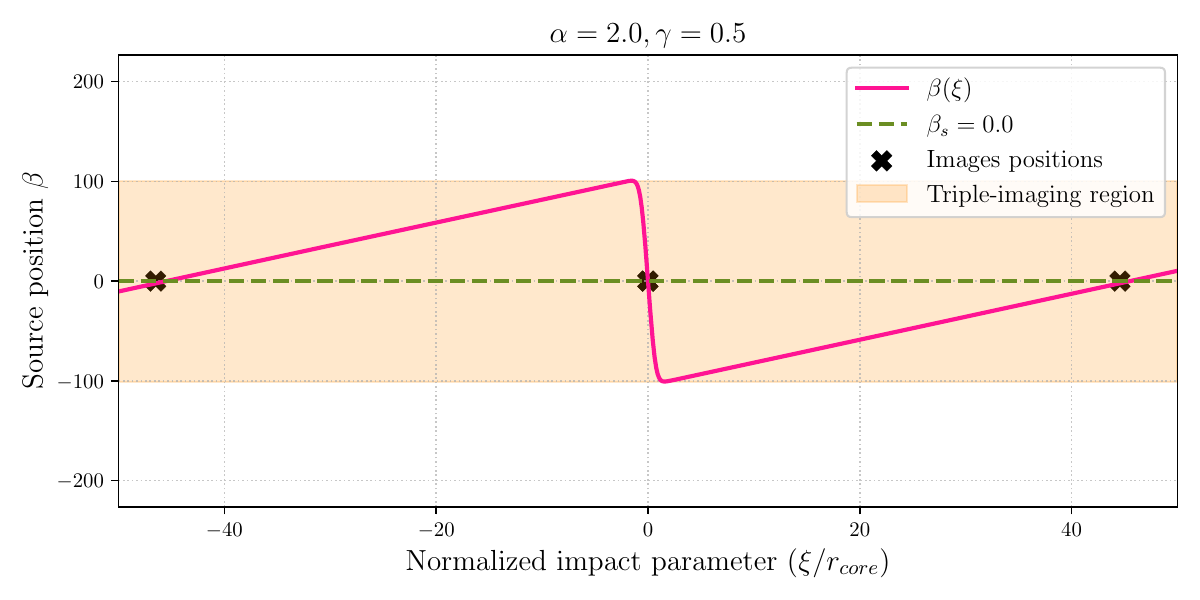}
    \caption{Representative lens mapping $\beta(\xi)$ for $\alpha=2.0$ and $\gamma=0.5$. Image positions are obtained from the solutions of $\beta(\xi)=\beta_s$, where $\beta_s$ denotes the source position. In this example, $\beta_s=0$.}
    \label{fig:beta_onecase}
\end{figure}

Figure~\ref{fig:beta_onecase} shows a representative example of the lens mapping $\beta(\xi)$. Within the interval $[-\beta_{\mathrm{lim}},+\beta_{\mathrm{lim}}]$, a given projected source position $\beta$ corresponds to three distinct values of the impact parameter $\xi$. According to the lens equation~\eqref{eq:lens_equation}, three distinct light rays connect the source and the observer, giving rise to triple imaging.

The intersections of the curve $\beta(\xi)$ with the horizontal axis correspond to image positions in the lens plane and are equivalently interpretable as measures of the angular separation between images. When the source is located at $\beta_s = 0$, the resulting images are symmetrically distributed, since the curve $\beta(\xi)$ intersects the line $\beta = \beta_s$ at symmetric values of $\xi$ with respect to the origin. In contrast, if $|\beta_s| < \beta_{\text{lim}}$ and $\beta_s \neq 0$, the intersection points between the lines $\beta = \beta_s$ and $\beta(\xi)$ are no longer symmetric with respect to $\xi = 0$, and therefore the three images are unevenly spaced.

\begin{figure}[h]
    \centering
    \includegraphics[width=0.75\linewidth]{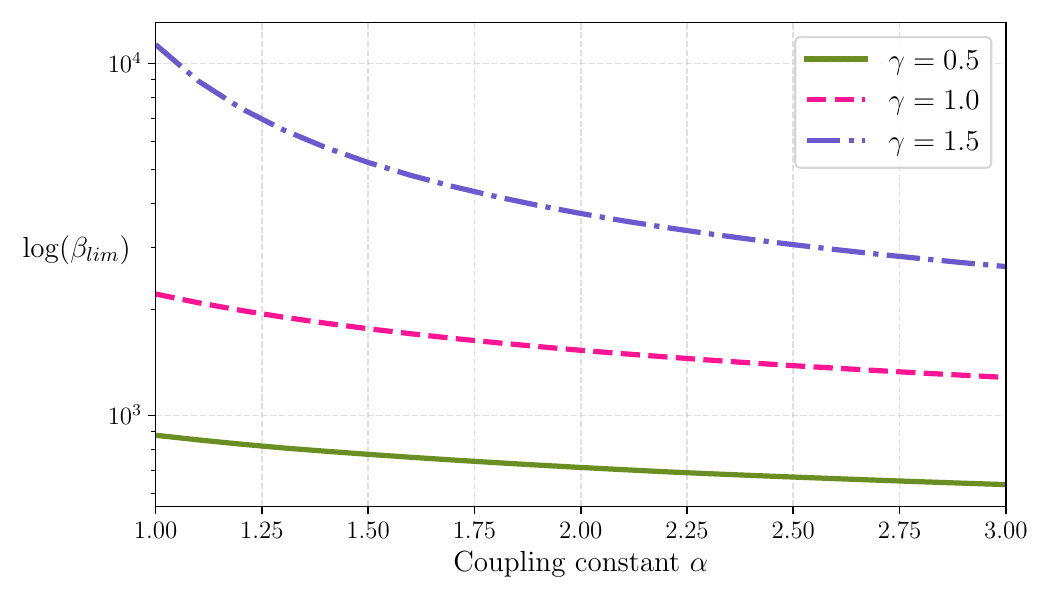}
    \caption{Dependence of $\log(\beta_{\mathrm{lim}})$ on the coupling parameter $\alpha$ for different values of the symmetry-breaking scale $\gamma$. Larger values of $\gamma$ (more massive strings) increase the range of source positions for which triple imaging occurs.}
    \label{fig:beta_limit}
\end{figure}

The quantity $\beta_{\mathrm{lim}}$ defines the largest displacement of the source from perfect alignment ($\beta_s=0$) for which three distinct image positions still exist, although not necessarily with symmetric image separations. Consequently, the interval
$[-\beta_{\mathrm{lim}}<\beta_s<\beta_{\mathrm{lim}}]$
determines the region in which triple imaging occurs.
In Figure \ref{fig:beta_limit} we show how the vortex parameters, $(\alpha,\gamma)$, influence the measure of $\beta_{\text{lim}}$. We observe that the triple-imaging configuration is mainly dictated by the symmetry-breaking scale $\gamma$, hence proportional to $|R(0)|$.

At this stage, the lens equation establishes the existence of three images but does not distinguish between their physical properties. Nevertheless, the distinct geodesic paths followed by the photons suggest that the images should carry different observational signatures.
Moreover, for perfectly aligned sources ($\beta_s = 0$), the external images form beyond the string's core ($\xi > r_{\text{core}}$) and thus encode the topological properties of the conical exterior determined by the deficit angle. 
Consequently, it is not unexpected that the angular separation between the exterior images is governed by the same expression as that obtained for the angular separation in the idealized case, namely $\Delta \varphi = 2 \pi (1 - b)/b$ \cite{vilenkin1981gravitational, gott1985gravitational, hiscock1985exact, linet1985static}. By contrast, the internal image traverses the strongly curved region near the vortex core and is consequently sensitive to the detailed structure of the Abelian-Higgs solution. The image magnification provides the distinction we are pursuing.

\subsection{Demagnification of the central image}

According to Liouville's theorem, in the absence of emission or absorption, the surface brightness of a source is conserved along null geodesics. It means that any variation in the observed total flux is due solely to a change in the apparent angular size of the image. Gravitational magnification is thus entirely determined by the geometry of the lens mapping.

\begin{figure}
    \centering
    \includegraphics[width=0.8\linewidth]{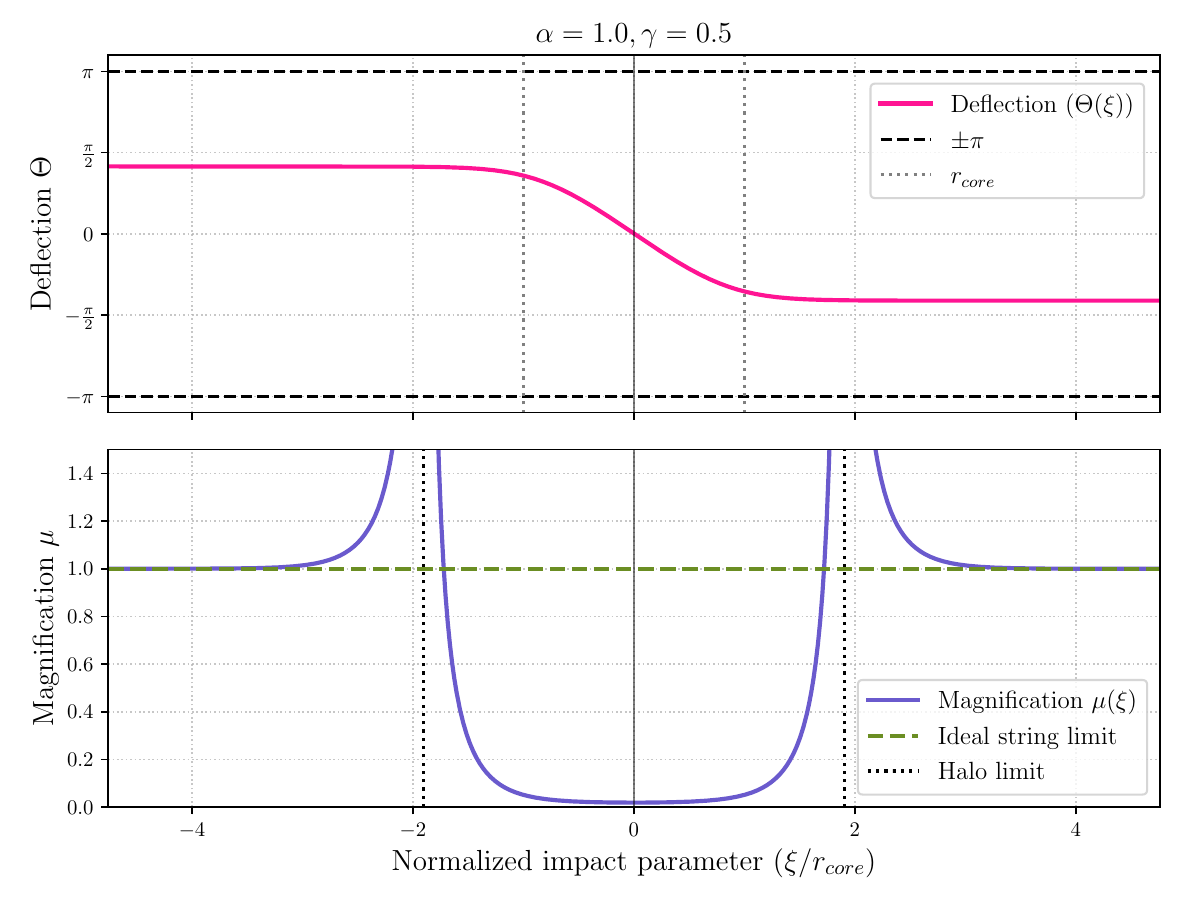}
    \caption{Deflection field $\Theta$ (top) and magnification of light rays $\mu(\xi)$ (bottom) as function of the normalized impact parameter $(\xi/r_{\text{core}})$. When $|\xi| \to \infty $ the deflection field is constant, meaning the photon deflection is purely topological, which gives a magnification $\mu = 1$.}
    \label{fig:mag_and_deflection}
\end{figure}

In our setup, the magnification is defined with the following Jacobian  \cite{meneghetti2021introduction},
\begin{equation}
\mu = \left| \frac{d\alpha}{d \tilde{\beta}} \right|,
\end{equation}
where $\alpha=-\Theta$ is the lensing deflection angle and $\tilde{\beta}$ is the angular source position introduced previously.
Using the relations $\tilde{\beta} = \beta/D_L$ together with $ \Theta = b/D_L$, one can connect $\mu$ and $\xi$ via
\begin{equation}
    \mu(\xi) = \left| \frac{\frac{1}{D_L} d\xi}{\frac{1}{D_L} d\beta} \right| = \left| \frac{d\beta}{d\xi} \right|^{-1}.
    \label{eq:magnification_raw}
\end{equation}
Now, notice that \eqref{eq:magnification_raw} can be calculated using the lens equation \eqref{eq:lens_equation}
\begin{equation}
    \mu(\xi) = \frac{1}{\left|1 + D_{\text{eff}} \Theta'(\xi) \right|},
    \label{eq:magnification}
\end{equation}
which connects the magnification with the derivative of deflection field with respect to the impact parameter, $\Theta'(\xi)$.
For images formed outside the core, where the deflection field is approximately constant ($\Theta(\xi \gg r_{\text{core}}) = \Theta_0$), the magnification approaches unity ($\mu(\xi \gg r_{\text{core}}) = 1$), guaranteeing that the external images have their intrinsic brightness conserved. Conversely, near the string's axis ($\xi \approx 0$), the gradient of the deflection angle, $|\Theta'(\xi)|$, reaches its maximum. This strong gradient induces a severe demagnification of the central image. Physically, this flux reduction arises from an increase in the proper cross-sectional area of the light beam, which dilutes the photon density. This areal expansion is driven by the convergence of light rays traversing the string's core, causing the exiting beam to diverge sharply.
Furthermore, since $|D_{\text{eff}}\Theta'(\xi)|$ is a continuous function ranging from $0$ to a peak at $\xi = 0$ (see figure~\ref{fig:mag_and_deflection}), there exist two critical points, $\pm \xi_{\text{halo}}$, where the inverse magnification $\mu^{-1}(\xi)$ vanishes, leading to a theoretical divergence in the magnification, $\mu(\pm \xi_{\text{halo}}) \to \infty$. These points define the luminous halo region $ |x| \leq \xi_{\text{halo}}$, often called the Einstein ring. At $\xi = \pm \xi_{\text{halo}}$ the proper area of the light bundle collapses to zero, representing a standard singularity within the geometrical optics formalism. Notably, the radial extent of this halo exceeds the core radius ($|\xi_{\text{halo}}| > r_{\text{core}}$), demonstrating that the string is optically thicker than its physical curvature well.

\begin{figure}
    \centering
    \includegraphics[width=0.65\linewidth]{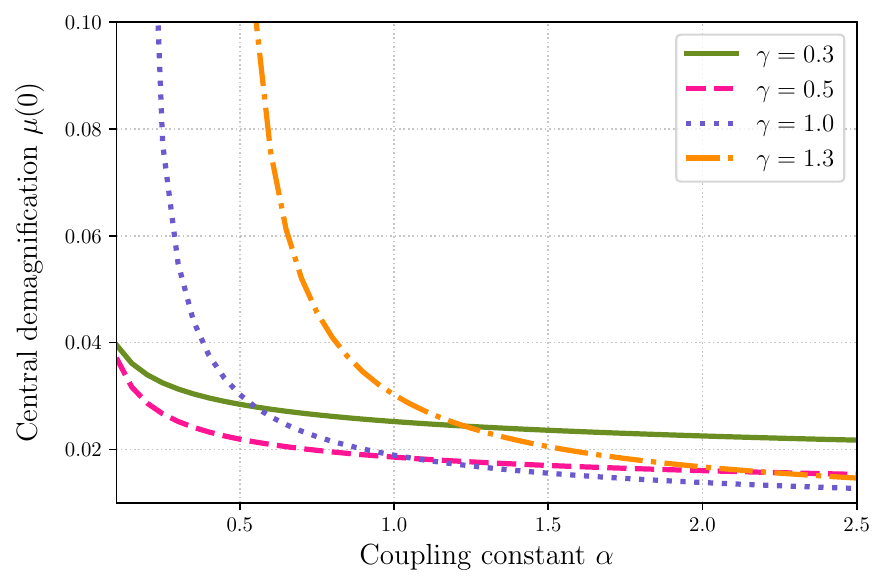}
    \caption{Demagnification of the central image $\mu(0)$. As we approach $\alpha \to \infty$, magnification strongly decays for all values of $\gamma$. As we reach the extremal limit, $\alpha \to 0$, the image becomes brighter due to the increasing vortex width $r_{\text{core}}$.}
\label{fig:mag_center_vs_alpha_vs_gamma}
\end{figure}

As illustrated in Figure~\ref{fig:mag_and_deflection}, the magnification of the central image, while small, does not identically vanish. Instead, it exhibits a complex dependence on the vortex parameters, which is further detailed in Figure~\ref{fig:mag_center_vs_alpha_vs_gamma}. Notice that magnification tends to zero as we approach the ideal string limit $\alpha \to \infty, \gamma \to \infty$ and is strongly enhanced when $\alpha \to 0$. Due to the increasing vortex size as $\alpha \to 0$, the deflection field has a subtle change around $r = 0$, i.e., $\Theta'(0)$ is small. In contrast, as $\alpha$ increases, $r_{\text{core}} \to 0$, squeezing $\Theta(\xi)$ and increasing $\Theta'(0)$ which diminishes magnification. In summary, when $\Theta'(0)$ is small, the photon experiences a slight change in the background geometry instead of a rigid barrier, which reduces the string's demagnifying power. 
Consequently, the string's gravitational potential acts as an effectively opaque lens, whose demagnifying power is regulated by the vortex radius. Furthermore, given that for each $\gamma$ there is a minimum allowed value of $\alpha$, defined by $\gamma = \gamma_{\text{crit}}(\alpha)$, we conclude that near-critical strings represent the most optically transparent configurations, as shown in Fig. \ref{fig:mag_center_vs_alpha_vs_gamma}.

\subsection{Shapiro time delay}

The presence of a metric component $g_{tt} = N^2(r)$ introduces a local time delay, commonly referred to as the Shapiro time delay. Unlike the ideal string studied in \cite{anderson2015mathematical, vilenkin1981gravitational, gott1985gravitational,hiscock1985exact, linet1985static}, where $N(r)=1$ everywhere and time delays are purely geometric, the internal structure of the gravitating string directly affects the coordinate time of light propagation.

To isolate the intrinsic effect of the vortex core from the asymptotic geometry, we define the Shapiro time delay, $\Delta T$, as the difference between the coordinate time elapsed in the gravitating string and the time that would be elapsed in an ideal conical background with a constant metric $N_{\infty} = a$. This measure ensures that we consider only the \emph{local} time delay due to the presence of the gravitating string, separating it from the delay arising from different cosmological paths. The cosmological observable, the \emph{total} time delay, is naturally a function of the path length of each image, which depends on the distances $D_i$.  

For a photon path parameterized by an affine parameter $\lambda$ with energy $E = -p_{t} = -g_{tt}\frac{dt}{d\lambda}$, the evolution of the coordinate time in the gravitating string spacetime is given by
\begin{equation} 
    dt_{\text{real}} = \frac{E}{N^2(r)} d\lambda_{\text{real}}.
    \label{eq:dt_real}
\end{equation}

Simultaneously, the null path condition, $N^2(r) dt_{\text{real}}^2 = dl_{\text{real}}^2$, relates the distance traveled by the photon, $dl_{\text{real}}$, to the affine parameter $\lambda$ via
\begin{equation} \label{eq:dl_real}
    dl_\text{real} = N(r) dt_\text{real}= \frac{E}{N(r)} d\lambda_{\text{real}}.
\end{equation}

Now, for a photon in the ideal string background, parameterized by its own affine parameter $\lambda_{\text{ideal}}$, the coordinate time evolution and the spatial distance traversed are
\begin{equation} \label{eq:dt_dl_ideal}
    dt_{\text{ideal}} = \frac{E}{a^2} d\lambda_{\text{ideal}}, \quad  dl_{\text{ideal}} = \frac{E}{a} d\lambda_{\text{ideal}}.
\end{equation}

To compute a physically meaningful time delay, we must compare the time taken by both photons to cover the same spatial distance, implying we must enforce $dl_{\text{real}} = dl_{\text{ideal}} \equiv dl$. Equating (\ref{eq:dl_real}) and (\ref{eq:dt_dl_ideal}), we find the transformation rule between the affine parameters of the two scenarios
\begin{equation} \label{eq:matching}
    \frac{E}{N(r)} d\lambda_\text{real} = \frac{E}{a} d\lambda_{\text{ideal}} \quad \implies \quad d\lambda_{\text{ideal}} = \frac{a}{N(r)} d\lambda_\text{real}.
\end{equation}

By substituting the matching condition \eqref{eq:matching} into the ideal time equation \eqref{eq:dt_dl_ideal}, and dropping the subindex ``real", we can express the reference time strictly in terms of the affine parameter $\lambda$ as follows
\begin{equation}
    dt_{\text{ideal}} = \frac{E}{a N(r)} d\lambda.
\end{equation}

Finally, the differential Shapiro time delay $d\tau_{S}$ is the difference between the coordinate time of the real vortex/string and the ideal string, which results in
\begin{equation}
    d\tau_{S} = dt_{\text{real}} - dt_{\text{ideal}} = E \left( \frac{1}{N^2(r)} - \frac{1}{a N(r)} \right) d\lambda.
    \label{eq:shapiro_time_delay}
\end{equation}

\begin{figure}
    \centering
    \includegraphics[width=0.75\linewidth]{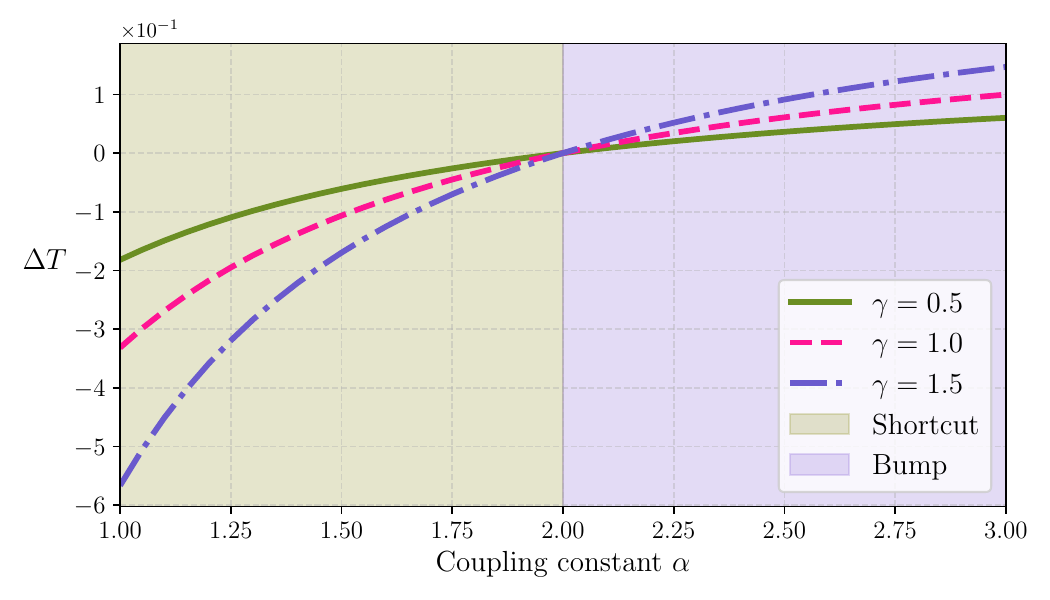}
    \caption{Difference between the time delay of the central ($C$) and the external ($A$) images, $\Delta T = \tau_C - \tau_A$. The central path can act as a shortcut or a bump depending on the value of $a$, which is regulated by the coupling constant $\alpha$.}
    \label{fig:time_delay}
\end{figure}

Integrating Eq.~\eqref{eq:shapiro_time_delay} along a photon trajectory yields the total Shapiro time delay associated with the string core. Since the integrand vanishes identically in the asymptotic region, where $N(r)\rightarrow a$, the resulting delay receives contributions only from the local curvature within ($r \lesssim r_{\text{core}}$) in which the metric deviates from its vacuum form. 
We are mainly interested in the time delay between signals, the difference in the accumulated time \eqref{eq:shapiro_time_delay} of the external(A) image and the central image(C), $\Delta T = \tau_C - \tau_A$. 
Our numerical analysis reveals a distinct transition in the temporal behavior of the central image relative to the external images, parametrically controlled by the coupling constant $\alpha$. As shown in Fig.~\ref{fig:time_delay}, the line corresponding to $\alpha = 2.0$ serves as a critical boundary separating three qualitatively different regimes. 

For $\alpha < 2$, the asymptotic normalization satisfies $a < N(0) = 1$. As a result, photons traversing the vortex core accumulate less coordinate time than photons propagating through the exterior region, giving rise to a negative relative delay, $\Delta T < 0$. In this regime, the central trajectory serves as a temporal shortcut relative to the external images. Conversely, for $\alpha > 2$, the relation $a > 1$ reverses this behavior: photons crossing the core accumulate more coordinate time than those traveling outside the vortex, leading to a positive delay, $\Delta T > 0$. At the self-dual point, $\alpha = 2$, one has $a = 1$, and the relative delay vanishes identically, $\Delta T = 0$.
This transition highlights a direct connection between the temporal properties of lensed images and the internal structure of the gravitating string. In particular, the sign and magnitude of the relative time delay encode information about the parameter $\alpha$, suggesting that, in principle, time-delay measurements could provide a direct observational probe of the finite-size gravitating strings.

\section{Conclusions}
\label{sec:conclusion}

In this work, we investigate the optical properties of gravitating Abelian-Higgs cosmic strings and compare them with those of the idealized infinitely thin string. Unlike the ideal string, which is characterized solely by a deficit angle, gravitating Abelian-Higgs strings possess a finite core with a nontrivial internal geometry. 
We showed that the size of the vortex core and the curvature generated around it depend strongly on the parameter $\alpha$, which controls the vortex's internal structure, and on the symmetry-breaking scale $\gamma$. As both parameters increase, the vortex becomes narrower, and the curvature becomes increasingly concentrated, approaching the ideal string limit.

Our analysis shows that the finite size of the string has important consequences for gravitational lensing. Most importantly, we found that gravitating strings can produce three images of the same source: two external images associated with light rays passing outside the core and a central image generated by photons that cross the string interior. This triple-imaging effect is absent in the ideal string approximation and therefore represents a direct signature of the internal structure of the vortex. We also show that the range of source positions for which triple imaging occurs increases with $\gamma$, making the effect more likely to be observed for heavier strings.
The angular separation between the two external images matches the prediction of the ideal cosmic string model. For perfectly aligned configurations, the external images are symmetric with respect to the string axis, while source misalignment breaks this symmetry. We quantified the maximum source displacement compatible with triple imaging and found that the effect becomes more robust for larger $\gamma$ and smaller $\alpha$, corresponding to strings with deeper curvature wells and broader cores.

The central image exhibits optical properties qualitatively different from those of the external images. While the external images preserve the source brightness, the central image is strongly demagnified. The degree of demagnification is controlled by the vortex core size and becomes increasingly pronounced as the string approaches the ideal limit. In this sense, the ideal cosmic string can be viewed as the limiting case in which the central image becomes observationally inaccessible.

A second distinctive feature of gravitating strings is the presence of a nontrivial Shapiro time delay. We found that the sign of the delay is controlled by the coupling constant $\alpha$. For $\alpha<2$, photons crossing the string core arrive earlier than those propagating outside the string, so that the core acts as a temporal shortcut. For $\alpha>2$, the situation is reversed, and the core behaves as a temporal barrier, producing a positive delay. At the self-dual point $\alpha=2$, the delay vanishes identically.

These results demonstrate that finite-width cosmic strings exhibit observational signatures entirely absent in the idealized Nambu-Goto description. The multiplicity of images, the demagnification of the central image, and the sign of the time delay all carry information about the vortex parameters of the underlying field theory. 
Consequently, if cosmic strings are ever observed through their lensing effects, these signatures may provide a direct way to probe the defect's internal structure and the physics responsible for its formation.

\appendix
\section{Numerical Approach}
\label{appendix:numerical appendix}The numerical method employed to solve the gravitating Abelian-Higgs background system \eqref{eq:field1}–\eqref{eq:field4} is organized as follows:

\begin{enumerate}
    \item The radial coordinate is discretized by introducing an $N$-point grid \texttt{x} together with the corresponding spectral differentiation matrix \texttt{D}. All derivatives are evaluated using \texttt{D}, and we adopt the notation \texttt{df} to denote the derivative of a grid function \texttt{f} with respect to \texttt{x}.
    \item The system \eqref{eq:field1}–\eqref{eq:field4} is assembled into a single residual vector \texttt{res(P,dP,f,df,N,dN,L,dL)}, with the appropriate boundary/initial conditions imposed. In this way, the full system may be written compactly as \texttt{EQ = 0}.
    \item The nonlinear system is then solved by finding a root of \texttt{res}, i.e., by determining the set of grid functions \texttt{[P,dP,f,df,N,dN,L,dL]} for which the right-hand sides of \eqref{eq:field1}–\eqref{eq:field4} vanish at all grid points.
\end{enumerate}

In step 1, a non-uniform Chebyshev grid is introduced in order to concentrate grid points near the origin. This is achieved via an algebraic mapping between the Chebyshev domain $[-1,1]$ and the physical domain $[0, r_\infty]$, where $r_\infty$ denotes the truncated asymptotic value of the radial coordinate. The matrix differentiation operator $D$ is then constructed such that, when it is multiplied by an array of function values, it yields the corresponding discrete derivative with respect to $x$. Our numerical construction is heavily based in \cite{trefethen2000spectral}. The adoption of Chebyshev spectral differentiation is motivated by its rapid convergence properties, typically requiring significantly smaller $N$ than high-order finite-difference schemes. This feature is particularly advantageous for the construction of the global residual in step 2.

In step 2, a residual function \texttt{res} is defined, which returns the left-hand sides of the system of equations assembled into a single $8N$-dimensional array, representing the local truncation error for each equation at each grid point in \texttt{x}. For each second-order differential equation in the system, an auxiliary first-order consistency equation is introduced to enforce compatibility between the discrete representation of the function and its derivative for $(P, f, N, L)$. Consequently, each entry of the array returned by \texttt{res} corresponds to the left-hand side of one of the equations evaluated at a specific grid point. Boundary conditions are enforced by prescribing the values of the boundary entries of the corresponding $N$-dimensional equation arrays.

In step 3, the function \texttt{res} is supplied to a multidimensional root-finding algorithm, which returns an $8N$-dimensional vector containing the numerical approximations to \texttt{[P,dP,f,df,N,dN,L,dL]}. This procedure was implemented in \texttt{Python}, employing the \texttt{root} method from the \texttt{scipy.minimize} class. The final numerical solution is characterized by a maximum absolute residual of order $10^{-10}$, as measured by the largest absolute component of the residual vector.

For the numerical integration of the photon equations of motion \eqref{eq:hamilton_eqs}, we employed the JAX-based solver \texttt{diffeqsolve} from the \texttt{diffrax} library, utilizing the fifth-order Runge–Kutta scheme \texttt{diffrax.Tsit5} with an adaptive step-size controller.

\begin{acknowledgments}
Both authors would like to thank CAPES (Coordenação de Aperfeiçoamento de Pessoal de Nível Superior) for funding. AM also acknowledges financial support from CNPq (Conselho Nacional de Desenvolvimento Científico e Tecnológico), Grant No.~306295/2023-7.
\end{acknowledgments}

\bibliography{references}

\begin{thebibliography}{32}%
\makeatletter
\providecommand \@ifxundefined [1]{%
 \@ifx{#1\undefined}
}%
\providecommand \@ifnum [1]{%
 \ifnum #1\expandafter \@firstoftwo
 \else \expandafter \@secondoftwo
 \fi
}%
\providecommand \@ifx [1]{%
 \ifx #1\expandafter \@firstoftwo
 \else \expandafter \@secondoftwo
 \fi
}%
\providecommand \natexlab [1]{#1}%
\providecommand \enquote  [1]{``#1''}%
\providecommand \bibnamefont  [1]{#1}%
\providecommand \bibfnamefont [1]{#1}%
\providecommand \citenamefont [1]{#1}%
\providecommand \href@noop [0]{\@secondoftwo}%
\providecommand \href [0]{\begingroup \@sanitize@url \@href}%
\providecommand \@href[1]{\@@startlink{#1}\@@href}%
\providecommand \@@href[1]{\endgroup#1\@@endlink}%
\providecommand \@sanitize@url [0]{\catcode `\\12\catcode `\$12\catcode `\&12\catcode `\#12\catcode `\^12\catcode `\_12\catcode `\%12\relax}%
\providecommand \@@startlink[1]{}%
\providecommand \@@endlink[0]{}%
\providecommand \url  [0]{\begingroup\@sanitize@url \@url }%
\providecommand \@url [1]{\endgroup\@href {#1}{\urlprefix }}%
\providecommand \urlprefix  [0]{URL }%
\providecommand \Eprint [0]{\href }%
\providecommand \doibase [0]{https://doi.org/}%
\providecommand \selectlanguage [0]{\@gobble}%
\providecommand \bibinfo  [0]{\@secondoftwo}%
\providecommand \bibfield  [0]{\@secondoftwo}%
\providecommand \translation [1]{[#1]}%
\providecommand \BibitemOpen [0]{}%
\providecommand \bibitemStop [0]{}%
\providecommand \bibitemNoStop [0]{.\EOS\space}%
\providecommand \EOS [0]{\spacefactor3000\relax}%
\providecommand \BibitemShut  [1]{\csname bibitem#1\endcsname}%
\let\auto@bib@innerbib\@empty
\bibitem [{\citenamefont {Kibble}(1976)}]{kibble1976topology}%
  \BibitemOpen
  \bibfield  {author} {\bibinfo {author} {\bibfnamefont {T.~W.}\ \bibnamefont {Kibble}},\ }\bibfield  {title} {\bibinfo {title} {Topology of cosmic domains and strings},\ }\href@noop {} {\bibfield  {journal} {\bibinfo  {journal} {Journal of Physics A: Mathematical and General}\ }\textbf {\bibinfo {volume} {9}},\ \bibinfo {pages} {1387} (\bibinfo {year} {1976})}\BibitemShut {NoStop}%
\bibitem [{\citenamefont {Kibble}(1980)}]{kibble1980some}%
  \BibitemOpen
  \bibfield  {author} {\bibinfo {author} {\bibfnamefont {T.~W.}\ \bibnamefont {Kibble}},\ }\bibfield  {title} {\bibinfo {title} {Some implications of a cosmological phase transition},\ }\href@noop {} {\bibfield  {journal} {\bibinfo  {journal} {Physics Reports}\ }\textbf {\bibinfo {volume} {67}},\ \bibinfo {pages} {183} (\bibinfo {year} {1980})}\BibitemShut {NoStop}%
\bibitem [{\citenamefont {Vilenkin}(1985)}]{vilenkin1985cosmic}%
  \BibitemOpen
  \bibfield  {author} {\bibinfo {author} {\bibfnamefont {A.}~\bibnamefont {Vilenkin}},\ }\bibfield  {title} {\bibinfo {title} {Cosmic strings and domain walls},\ }\href@noop {} {\bibfield  {journal} {\bibinfo  {journal} {Physics reports}\ }\textbf {\bibinfo {volume} {121}},\ \bibinfo {pages} {263} (\bibinfo {year} {1985})}\BibitemShut {NoStop}%
\bibitem [{\citenamefont {Vilenkin}\ \emph {et~al.}(1994)\citenamefont {Vilenkin}, \citenamefont {Vilenkin},\ and\ \citenamefont {Shellard}}]{vilenkin1994cosmic}%
  \BibitemOpen
  \bibfield  {author} {\bibinfo {author} {\bibfnamefont {A.}~\bibnamefont {Vilenkin}}, \bibinfo {author} {\bibfnamefont {A.}~\bibnamefont {Vilenkin}},\ and\ \bibinfo {author} {\bibfnamefont {E.}~\bibnamefont {Shellard}},\ }\href@noop {} {\emph {\bibinfo {title} {Cosmic strings and other topological defects}}}\ (\bibinfo  {publisher} {Cambridge University Press},\ \bibinfo {year} {1994})\BibitemShut {NoStop}%
\bibitem [{\citenamefont {Ferreira}(2021)}]{ferreira2021ultra}%
  \BibitemOpen
  \bibfield  {author} {\bibinfo {author} {\bibfnamefont {E.~G.}\ \bibnamefont {Ferreira}},\ }\bibfield  {title} {\bibinfo {title} {Ultra-light dark matter},\ }\href@noop {} {\bibfield  {journal} {\bibinfo  {journal} {The Astronomy and Astrophysics Review}\ }\textbf {\bibinfo {volume} {29}},\ \bibinfo {pages} {7} (\bibinfo {year} {2021})}\BibitemShut {NoStop}%
\bibitem [{\citenamefont {Schive}\ \emph {et~al.}(2014)\citenamefont {Schive}, \citenamefont {Chiueh},\ and\ \citenamefont {Broadhurst}}]{schive2014cosmic}%
  \BibitemOpen
  \bibfield  {author} {\bibinfo {author} {\bibfnamefont {H.-Y.}\ \bibnamefont {Schive}}, \bibinfo {author} {\bibfnamefont {T.}~\bibnamefont {Chiueh}},\ and\ \bibinfo {author} {\bibfnamefont {T.}~\bibnamefont {Broadhurst}},\ }\bibfield  {title} {\bibinfo {title} {Cosmic structure as the quantum interference of a coherent dark wave},\ }\href@noop {} {\bibfield  {journal} {\bibinfo  {journal} {Nature Physics}\ }\textbf {\bibinfo {volume} {10}},\ \bibinfo {pages} {496} (\bibinfo {year} {2014})}\BibitemShut {NoStop}%
\bibitem [{\citenamefont {Hu}\ \emph {et~al.}(2000)\citenamefont {Hu}, \citenamefont {Barkana},\ and\ \citenamefont {Gruzinov}}]{hu2000fuzzy}%
  \BibitemOpen
  \bibfield  {author} {\bibinfo {author} {\bibfnamefont {W.}~\bibnamefont {Hu}}, \bibinfo {author} {\bibfnamefont {R.}~\bibnamefont {Barkana}},\ and\ \bibinfo {author} {\bibfnamefont {A.}~\bibnamefont {Gruzinov}},\ }\bibfield  {title} {\bibinfo {title} {Fuzzy cold dark matter: the wave properties of ultralight particles},\ }\href@noop {} {\bibfield  {journal} {\bibinfo  {journal} {Physical Review Letters}\ }\textbf {\bibinfo {volume} {85}},\ \bibinfo {pages} {1158} (\bibinfo {year} {2000})}\BibitemShut {NoStop}%
\bibitem [{\citenamefont {Desjacques}\ \emph {et~al.}(2018)\citenamefont {Desjacques}, \citenamefont {Kehagias},\ and\ \citenamefont {Riotto}}]{desjacques2018impact}%
  \BibitemOpen
  \bibfield  {author} {\bibinfo {author} {\bibfnamefont {V.}~\bibnamefont {Desjacques}}, \bibinfo {author} {\bibfnamefont {A.}~\bibnamefont {Kehagias}},\ and\ \bibinfo {author} {\bibfnamefont {A.}~\bibnamefont {Riotto}},\ }\bibfield  {title} {\bibinfo {title} {Impact of ultralight axion self-interactions on the large scale structure of the universe},\ }\href@noop {} {\bibfield  {journal} {\bibinfo  {journal} {Physical Review D}\ }\textbf {\bibinfo {volume} {97}},\ \bibinfo {pages} {023529} (\bibinfo {year} {2018})}\BibitemShut {NoStop}%
\bibitem [{\citenamefont {Brax}\ and\ \citenamefont {Valageas}(2025)}]{brax20253d}%
  \BibitemOpen
  \bibfield  {author} {\bibinfo {author} {\bibfnamefont {P.}~\bibnamefont {Brax}}\ and\ \bibinfo {author} {\bibfnamefont {P.}~\bibnamefont {Valageas}},\ }\bibfield  {title} {\bibinfo {title} {3d vortices and rotating solitons in ultralight dark matter},\ }\href@noop {} {\bibfield  {journal} {\bibinfo  {journal} {Physical Review D}\ }\textbf {\bibinfo {volume} {111}},\ \bibinfo {pages} {103538} (\bibinfo {year} {2025})}\BibitemShut {NoStop}%
\bibitem [{\citenamefont {Rindler-Daller}\ and\ \citenamefont {Shapiro}(2012)}]{rindler2012angular}%
  \BibitemOpen
  \bibfield  {author} {\bibinfo {author} {\bibfnamefont {T.}~\bibnamefont {Rindler-Daller}}\ and\ \bibinfo {author} {\bibfnamefont {P.~R.}\ \bibnamefont {Shapiro}},\ }\bibfield  {title} {\bibinfo {title} {Angular momentum and vortex formation in bose--einstein-condensed cold dark matter haloes},\ }\href@noop {} {\bibfield  {journal} {\bibinfo  {journal} {Monthly Notices of the Royal Astronomical Society}\ }\textbf {\bibinfo {volume} {422}},\ \bibinfo {pages} {135} (\bibinfo {year} {2012})}\BibitemShut {NoStop}%
\bibitem [{\citenamefont {Kamionkowski}\ and\ \citenamefont {March-Russell}(1992)}]{kamionkowski1992planck}%
  \BibitemOpen
  \bibfield  {author} {\bibinfo {author} {\bibfnamefont {M.}~\bibnamefont {Kamionkowski}}\ and\ \bibinfo {author} {\bibfnamefont {J.}~\bibnamefont {March-Russell}},\ }\bibfield  {title} {\bibinfo {title} {Planck-scale physics and the peccei-quinn mechanism},\ }\href@noop {} {\bibfield  {journal} {\bibinfo  {journal} {Physics Letters B}\ }\textbf {\bibinfo {volume} {282}},\ \bibinfo {pages} {137} (\bibinfo {year} {1992})}\BibitemShut {NoStop}%
\bibitem [{\citenamefont {Vilenkin}(1981)}]{vilenkin1981gravitational}%
  \BibitemOpen
  \bibfield  {author} {\bibinfo {author} {\bibfnamefont {A.}~\bibnamefont {Vilenkin}},\ }\bibfield  {title} {\bibinfo {title} {Gravitational field of vacuum domain walls and strings},\ }\href@noop {} {\bibfield  {journal} {\bibinfo  {journal} {Physical Review D}\ }\textbf {\bibinfo {volume} {23}},\ \bibinfo {pages} {852} (\bibinfo {year} {1981})}\BibitemShut {NoStop}%
\bibitem [{\citenamefont {Gott~III}(1985)}]{gott1985gravitational}%
  \BibitemOpen
  \bibfield  {author} {\bibinfo {author} {\bibfnamefont {J.~R.}\ \bibnamefont {Gott~III}},\ }\bibfield  {title} {\bibinfo {title} {Gravitational lensing effects of vacuum strings-exact solutions},\ }\href@noop {} {\bibfield  {journal} {\bibinfo  {journal} {Astrophysical Journal, Part 1 (ISSN 0004-637X), vol. 288, Jan. 15, 1985, p. 422-427.}\ }\textbf {\bibinfo {volume} {288}},\ \bibinfo {pages} {422} (\bibinfo {year} {1985})}\BibitemShut {NoStop}%
\bibitem [{\citenamefont {Hiscock}(1985)}]{hiscock1985exact}%
  \BibitemOpen
  \bibfield  {author} {\bibinfo {author} {\bibfnamefont {W.~A.}\ \bibnamefont {Hiscock}},\ }\bibfield  {title} {\bibinfo {title} {Exact gravitational field of a string},\ }\href@noop {} {\bibfield  {journal} {\bibinfo  {journal} {Physical Review D}\ }\textbf {\bibinfo {volume} {31}},\ \bibinfo {pages} {3288} (\bibinfo {year} {1985})}\BibitemShut {NoStop}%
\bibitem [{\citenamefont {Bulygin}\ \emph {et~al.}(2023)\citenamefont {Bulygin}, \citenamefont {Sazhin},\ and\ \citenamefont {Sazhina}}]{bulygin2023theory}%
  \BibitemOpen
  \bibfield  {author} {\bibinfo {author} {\bibfnamefont {I.~I.}\ \bibnamefont {Bulygin}}, \bibinfo {author} {\bibfnamefont {M.~V.}\ \bibnamefont {Sazhin}},\ and\ \bibinfo {author} {\bibfnamefont {O.~S.}\ \bibnamefont {Sazhina}},\ }\bibfield  {title} {\bibinfo {title} {Theory of gravitational lensing on a curved cosmic string},\ }\href@noop {} {\bibfield  {journal} {\bibinfo  {journal} {The European Physical Journal C}\ }\textbf {\bibinfo {volume} {83}},\ \bibinfo {pages} {844} (\bibinfo {year} {2023})}\BibitemShut {NoStop}%
\bibitem [{\citenamefont {Fern{\'a}ndez-N{\'u}{\~n}ez}\ and\ \citenamefont {Bulashenko}(2017)}]{fernandez2017emergence}%
  \BibitemOpen
  \bibfield  {author} {\bibinfo {author} {\bibfnamefont {I.}~\bibnamefont {Fern{\'a}ndez-N{\'u}{\~n}ez}}\ and\ \bibinfo {author} {\bibfnamefont {O.}~\bibnamefont {Bulashenko}},\ }\bibfield  {title} {\bibinfo {title} {Emergence of fresnel diffraction zones in gravitational lensing by a cosmic string},\ }\href@noop {} {\bibfield  {journal} {\bibinfo  {journal} {Physics Letters A}\ }\textbf {\bibinfo {volume} {381}},\ \bibinfo {pages} {1764} (\bibinfo {year} {2017})}\BibitemShut {NoStop}%
\bibitem [{\citenamefont {Garfinkle}(1985)}]{garfinkle1985general}%
  \BibitemOpen
  \bibfield  {author} {\bibinfo {author} {\bibfnamefont {D.}~\bibnamefont {Garfinkle}},\ }\bibfield  {title} {\bibinfo {title} {General relativistic strings},\ }\href@noop {} {\bibfield  {journal} {\bibinfo  {journal} {Physical Review D}\ }\textbf {\bibinfo {volume} {32}},\ \bibinfo {pages} {1323} (\bibinfo {year} {1985})}\BibitemShut {NoStop}%
\bibitem [{\citenamefont {Linet}(1985)}]{linet1985static}%
  \BibitemOpen
  \bibfield  {author} {\bibinfo {author} {\bibfnamefont {B.}~\bibnamefont {Linet}},\ }\bibfield  {title} {\bibinfo {title} {The static metrics with cylindrical symmetry describing a model of cosmic strings},\ }\href@noop {} {\bibfield  {journal} {\bibinfo  {journal} {General Relativity and Gravitation}\ }\textbf {\bibinfo {volume} {17}},\ \bibinfo {pages} {1109} (\bibinfo {year} {1985})}\BibitemShut {NoStop}%
\bibitem [{\citenamefont {Christensen}\ \emph {et~al.}(1999)\citenamefont {Christensen}, \citenamefont {Larsen},\ and\ \citenamefont {Verbin}}]{christensen1999complete}%
  \BibitemOpen
  \bibfield  {author} {\bibinfo {author} {\bibfnamefont {M.}~\bibnamefont {Christensen}}, \bibinfo {author} {\bibfnamefont {A.}~\bibnamefont {Larsen}},\ and\ \bibinfo {author} {\bibfnamefont {Y.}~\bibnamefont {Verbin}},\ }\bibfield  {title} {\bibinfo {title} {Complete classification of the string-like solutions of the gravitating abelian higgs model},\ }\href@noop {} {\bibfield  {journal} {\bibinfo  {journal} {Physical Review D}\ }\textbf {\bibinfo {volume} {60}},\ \bibinfo {pages} {125012} (\bibinfo {year} {1999})}\BibitemShut {NoStop}%
\bibitem [{\citenamefont {Brihaye}\ and\ \citenamefont {Lubo}(2000)}]{brihaye2000classical}%
  \BibitemOpen
  \bibfield  {author} {\bibinfo {author} {\bibfnamefont {Y.}~\bibnamefont {Brihaye}}\ and\ \bibinfo {author} {\bibfnamefont {M.}~\bibnamefont {Lubo}},\ }\bibfield  {title} {\bibinfo {title} {Classical solutions of the gravitating abelian higgs model},\ }\href@noop {} {\bibfield  {journal} {\bibinfo  {journal} {Physical Review D}\ }\textbf {\bibinfo {volume} {62}},\ \bibinfo {pages} {085004} (\bibinfo {year} {2000})}\BibitemShut {NoStop}%
\bibitem [{\citenamefont {Hartmann}\ and\ \citenamefont {Sirimachan}(2010)}]{hartmann2010geodesic}%
  \BibitemOpen
  \bibfield  {author} {\bibinfo {author} {\bibfnamefont {B.}~\bibnamefont {Hartmann}}\ and\ \bibinfo {author} {\bibfnamefont {P.}~\bibnamefont {Sirimachan}},\ }\bibfield  {title} {\bibinfo {title} {Geodesic motion in the space-time of a cosmic string},\ }\href@noop {} {\bibfield  {journal} {\bibinfo  {journal} {Journal of High Energy Physics}\ }\textbf {\bibinfo {volume} {2010}},\ \bibinfo {pages} {1} (\bibinfo {year} {2010})}\BibitemShut {NoStop}%
\bibitem [{\citenamefont {Hartmann}\ \emph {et~al.}(2011)\citenamefont {Hartmann}, \citenamefont {L{\"a}mmerzahl},\ and\ \citenamefont {Sirimachan}}]{hartmann2011detection}%
  \BibitemOpen
  \bibfield  {author} {\bibinfo {author} {\bibfnamefont {B.}~\bibnamefont {Hartmann}}, \bibinfo {author} {\bibfnamefont {C.}~\bibnamefont {L{\"a}mmerzahl}},\ and\ \bibinfo {author} {\bibfnamefont {P.}~\bibnamefont {Sirimachan}},\ }\bibfield  {title} {\bibinfo {title} {Detection of cosmic superstrings by geodesic test particle motion},\ }\href@noop {} {\bibfield  {journal} {\bibinfo  {journal} {Physical Review D—Particles, Fields, Gravitation, and Cosmology}\ }\textbf {\bibinfo {volume} {83}},\ \bibinfo {pages} {045027} (\bibinfo {year} {2011})}\BibitemShut {NoStop}%
\bibitem [{\citenamefont {Hartmann}\ and\ \citenamefont {Kagramanova}(2012)}]{hartmann2012geodesic}%
  \BibitemOpen
  \bibfield  {author} {\bibinfo {author} {\bibfnamefont {B.}~\bibnamefont {Hartmann}}\ and\ \bibinfo {author} {\bibfnamefont {V.}~\bibnamefont {Kagramanova}},\ }\bibfield  {title} {\bibinfo {title} {Geodesic motion in the space-time of cosmic strings interacting via magnetic fields},\ }\href@noop {} {\bibfield  {journal} {\bibinfo  {journal} {Physical Review D—Particles, Fields, Gravitation, and Cosmology}\ }\textbf {\bibinfo {volume} {86}},\ \bibinfo {pages} {045028} (\bibinfo {year} {2012})}\BibitemShut {NoStop}%
\bibitem [{\citenamefont {Nielsen}\ and\ \citenamefont {Olesen}(1973)}]{nielsen1973vortex}%
  \BibitemOpen
  \bibfield  {author} {\bibinfo {author} {\bibfnamefont {H.~B.}\ \bibnamefont {Nielsen}}\ and\ \bibinfo {author} {\bibfnamefont {P.}~\bibnamefont {Olesen}},\ }\bibfield  {title} {\bibinfo {title} {Vortex-line models for dual strings},\ }\href@noop {} {\bibfield  {journal} {\bibinfo  {journal} {Nuclear Physics B}\ }\textbf {\bibinfo {volume} {61}},\ \bibinfo {pages} {45} (\bibinfo {year} {1973})}\BibitemShut {NoStop}%
\bibitem [{\citenamefont {Comtet}\ and\ \citenamefont {Gibbons}(1988)}]{comtet1988bogomol}%
  \BibitemOpen
  \bibfield  {author} {\bibinfo {author} {\bibfnamefont {A.}~\bibnamefont {Comtet}}\ and\ \bibinfo {author} {\bibfnamefont {G.}~\bibnamefont {Gibbons}},\ }\bibfield  {title} {\bibinfo {title} {Bogomol'nyi bounds for cosmic strings},\ }\href@noop {} {\bibfield  {journal} {\bibinfo  {journal} {Nuclear Physics B}\ }\textbf {\bibinfo {volume} {299}},\ \bibinfo {pages} {719} (\bibinfo {year} {1988})}\BibitemShut {NoStop}%
\bibitem [{\citenamefont {Linet}(1987)}]{linet1987vortex}%
  \BibitemOpen
  \bibfield  {author} {\bibinfo {author} {\bibfnamefont {B.}~\bibnamefont {Linet}},\ }\bibfield  {title} {\bibinfo {title} {A vortex-line model for infinite straight cosmic strings},\ }\href@noop {} {\bibfield  {journal} {\bibinfo  {journal} {Physics Letters A}\ }\textbf {\bibinfo {volume} {124}},\ \bibinfo {pages} {240} (\bibinfo {year} {1987})}\BibitemShut {NoStop}%
\bibitem [{\citenamefont {Geroch}\ and\ \citenamefont {Traschen}(1987)}]{geroch1987strings}%
  \BibitemOpen
  \bibfield  {author} {\bibinfo {author} {\bibfnamefont {R.}~\bibnamefont {Geroch}}\ and\ \bibinfo {author} {\bibfnamefont {J.}~\bibnamefont {Traschen}},\ }\bibfield  {title} {\bibinfo {title} {Strings and other distributional sources in general relativity},\ }\href@noop {} {\bibfield  {journal} {\bibinfo  {journal} {Physical Review D}\ }\textbf {\bibinfo {volume} {36}},\ \bibinfo {pages} {1017} (\bibinfo {year} {1987})}\BibitemShut {NoStop}%
\bibitem [{\citenamefont {Burke}(1981)}]{burke1981multiple}%
  \BibitemOpen
  \bibfield  {author} {\bibinfo {author} {\bibfnamefont {W.~L.}\ \bibnamefont {Burke}},\ }\bibfield  {title} {\bibinfo {title} {Multiple gravitational imaging by distributed masses},\ }\href@noop {} {\bibfield  {journal} {\bibinfo  {journal} {Astrophysical Journal, Vol. 244, P. L1, 1981}\ }\textbf {\bibinfo {volume} {244}},\ \bibinfo {pages} {L1} (\bibinfo {year} {1981})}\BibitemShut {NoStop}%
\bibitem [{\citenamefont {McKenzie}(1985)}]{mckenzie1985gravitational}%
  \BibitemOpen
  \bibfield  {author} {\bibinfo {author} {\bibfnamefont {R.~H.}\ \bibnamefont {McKenzie}},\ }\bibfield  {title} {\bibinfo {title} {A gravitational lens produces an odd number of images},\ }\href@noop {} {\bibfield  {journal} {\bibinfo  {journal} {Journal of mathematical physics}\ }\textbf {\bibinfo {volume} {26}},\ \bibinfo {pages} {1592} (\bibinfo {year} {1985})}\BibitemShut {NoStop}%
\bibitem [{\citenamefont {Meneghetti}(2021)}]{meneghetti2021introduction}%
  \BibitemOpen
  \bibfield  {author} {\bibinfo {author} {\bibfnamefont {M.}~\bibnamefont {Meneghetti}},\ }\href@noop {} {\emph {\bibinfo {title} {Introduction to gravitational lensing: with Python examples}}}\ (\bibinfo  {publisher} {Springer Nature},\ \bibinfo {year} {2021})\BibitemShut {NoStop}%
\bibitem [{\citenamefont {Anderson}(2015)}]{anderson2015mathematical}%
  \BibitemOpen
  \bibfield  {author} {\bibinfo {author} {\bibfnamefont {M.~R.}\ \bibnamefont {Anderson}},\ }\href@noop {} {\emph {\bibinfo {title} {The mathematical theory of cosmic strings: cosmic strings in the wire approximation}}}\ (\bibinfo  {publisher} {CRC Press},\ \bibinfo {year} {2015})\BibitemShut {NoStop}%
\bibitem [{\citenamefont {Trefethen}(2000)}]{trefethen2000spectral}%
  \BibitemOpen
  \bibfield  {author} {\bibinfo {author} {\bibfnamefont {L.~N.}\ \bibnamefont {Trefethen}},\ }\href@noop {} {\emph {\bibinfo {title} {Spectral methods in MATLAB}}}\ (\bibinfo  {publisher} {SIAM},\ \bibinfo {year} {2000})\BibitemShut {NoStop}%
\end{thebibliography}%
\end{document}